\documentclass{WileyMSP-template}

\usepackage{textcomp}

\usepackage{amsmath}
\usepackage{graphicx}
\usepackage[colorlinks=true, allcolors=blue]{hyperref}
\usepackage{amssymb}
\usepackage{multicol}
\usepackage{changepage}
\usepackage{gensymb}
\usepackage{parskip}
\usepackage{multirow}
\usepackage{ragged2e}
\usepackage{lineno}
\justifying

\usepackage{xr}
\begin{document}


\title{Fabric Phononic Crystals for Passive Vibration Control}

\maketitle


\author{Michael Y Wang$^{1}$}
\author{Hridyesh Tewani$^{2}$}
\author{Marianne Fairbanks$^{4}$}
\author{Pavana Prabhakar$^{2,3}$}
\author{Chu Ma$^{1,*}$}

\begin{affiliations}
\noindent
$^{1}$Department of Electrical and Computer Engineering, College of Engineering, University of Wisconsin–Madison, 1415 Engineering Drive, Madison, WI 53706, USA\\
$^{2}$Department of Civil and Environmental Engineering, College of Engineering, University of Wisconsin–Madison, 1415 Engineering Drive, Madison, WI 53706, USA\\
$^{3}$Department of Mechanical Engineering, College of Engineering, University of Wisconsin–Madison, 1513 University Avenue, Madison, WI 53706, USA\\
$^{4}$Department of Design Studies, School of Human Ecology, University of Wisconsin–Madison, 1300 Linden Dr, Madison, WI 53706, USA\\
\end{affiliations}

Email Address: chu.ma@wisc.edu


\begin{abstract}

Weaving patterns in fabrics, traditionally used for aesthetic purposes, present a largely untapped opportunity to create metamaterials that serve as passive layers for sensing, filtering, and signal processing. However, the hierarchical architecture of fabrics makes structural design and wave prediction challenging. Here, we establish fully woven fabrics as phononic crystals that passively filter and route elastic vibrations. Using double weaving, we integrate a soft cotton weave with stiff woven copper inclusions to form periodic fabric lattices with engineered dispersion. A multiscale modeling framework that combines homogenization of weave blocks with an effective-property macroscale model enables computationally efficient design of phononic crystals. Simulations and experiments confirm a pronounced phononic bandgap for out-of-plane vibrations in a finite fabric crystal, while an equivalent pure cotton weave shows no band suppression in the corresponding frequency range. Building on the same platform, we realize a fully woven higher-order topological insulator. Modal analysis and transmission measurements reveal in-gap edge states and localized corner states. These results show that phononic bandgaps and topological states can be directly encoded through weaving patterns and material contrast, enabling passive vibration‑management layers and multifunctional wave‑guiding fabrics for sensing, haptic interfaces, robotics, and noise mitigation.

\end{abstract}

\section{Introduction}

Fabrics have long been used as protective layers against wind, heat, cold, and injury, as well as for decorative purposes \cite{kornreich1966introduction, fan2009engineering}. High‑performance textiles further extend these roles to applications such as soft ballistic‑impact protection in bulletproof vests and turbine‑engine fragment barriers \cite{sockalingam2017recent}. Recently, fabrics have evolved into “smart” materials that integrate mechanical, acoustic, electrical, thermal, optical, and other advanced functions \cite{castano2014smart, loke2020recent, loke2021digital}. These capabilities are typically achieved by embedding sensors, actuators, and circuits directly into the textile structure. However, implementing complex sensing functions often requires large numbers of active components, which increases fabrication complexity, maintenance challenges, and power consumption \cite{takamatsu2012fabric, shu2010shoe, lee2019ultrathin, kim2019stretchable, catrysse2004towards, abouraddy2006large, lumelsky2000sensitive, someya2005conformable, hasegawa2007novel, meyer2010design, takamatsu2011meter, grant2004developing}. For example, an 8×8 pressure-sensor array was embedded into a 16 cm×16 cm fabric to sense the touch input for human-machine interface \cite{takamatsu2012fabric}. Similarly, a 4×5 acoustic-microphone array was produced on a large-area conformal textile substrate for directional sound detection \cite{grant2004developing, salvado2012intelligent}. 

Phononic crystals are periodically arranged wave scatterers that interact with acoustic/mechanical waves in a similar way to how atomic lattices interact with electronic waves. Waves of specific frequencies and momenta are allowed to propagate in the periodic system, leading to pass bands and bandgaps in the dispersion relations. In the bandgaps, wave propagation is prohibited along certain, or all, directions \cite{ma2013optimization, ge2018breaking, martinez1995sound, sanchez1998sound}. Within the band gap, waves can be trapped at a point defect, or propagate along a line or surface that can serve as a waveguide. Phononic crystals have been widely explored by researchers for their exotic properties when interacting with acoustic/elastic waves, such as negative refraction~\cite{martinez1995sound, sanchez1998sound, zhang2004negative, feng2005negative, feng2006acoustic}, acoustic collimation \cite{yang2004focusing, ke2005negative, qiu2005far, sukhovich2009experimental}, directional acoustic signal transmittance and reception~ \cite{ma2013optimization, he2016acoustic, zhang2019dimensional, zhang2019second}, and topologically robust and defect-insensitive wave guiding and splitting \cite{he2016acoustic, zhang2017topological, dai2019temperature, zhang2019dimensional, wang2022underwater, zheng2023switchable, zhang2019second}, all of which provides new opportunities for acoustic/mechanical wave transport and manipulation, information science, and sensing. Because fabrics are intrinsically hierarchical structures composed of yarns, their weaving patterns, traditionally used primarily for aesthetic purposes, present a largely untapped opportunity to create phononic crystals that can serve as passive layers for sensing, filtering, and signal processing. 

In this work, we introduce and experimentally validate a new class of fabric‑based phononic crystals, engineered by creating periodic patterns of fabric patches with different mechanical properties in woven fabrics. Our approach leverages the intrinsic geometric and material versatility of weaving to combine soft and light-weight cotton yarns with stiff and heavy metallic fibers, creating a large mechanical contrast that enables precise control over elastic wave propagation. By systematically tailoring this contrast and the periodic architecture, we designed fully woven fabric sheets that exhibit targeted phononic band structures capable of manipulating vibrations. To accurately predict and optimize the dynamic behavior of the fabrics, we developed a multiscale modeling framework that integrates detailed mesoscale simulations at the yarn level with homogenized macroscale models for full fabric analysis. This computational approach is validated through vibration transmission experiments conducted on fully woven fabric samples. Out-of-plane vibrations excited by a point shaker over the $1–100$~Hz range are captured using a  camera, enabling direct comparison with the simulated displacement fields. Both numerical and experimental results confirm the targeted band structures. Leveraging this design platform, we demonstrate two representative phononic fabric architectures for controlling elastic vibrations: (i) a bandstop filter exhibiting a bandgap that prohibits wave propagation, and (ii) a topological insulator that confines vibrations to selected edges and corners. Our fully woven, entirely passive phononic crystals opens new opportunities for vibration isolation, wearable acoustic and motion sensing, and lightweight noise‑mitigating structures. Because the vibration‑manipulation capabilities arise passively from the fabric architecture itself, these functions substantially reduce system complexity and power consumption compared with existing smart‑fabric strategies for vibration sensing and control, which typically depend on dense arrays of active electronic components.

\section {Results and Discussion}

\subsection{Design and multiscale modeling}

\begin{figure}[ht!]
\centering
\includegraphics[width=0.9\textwidth]{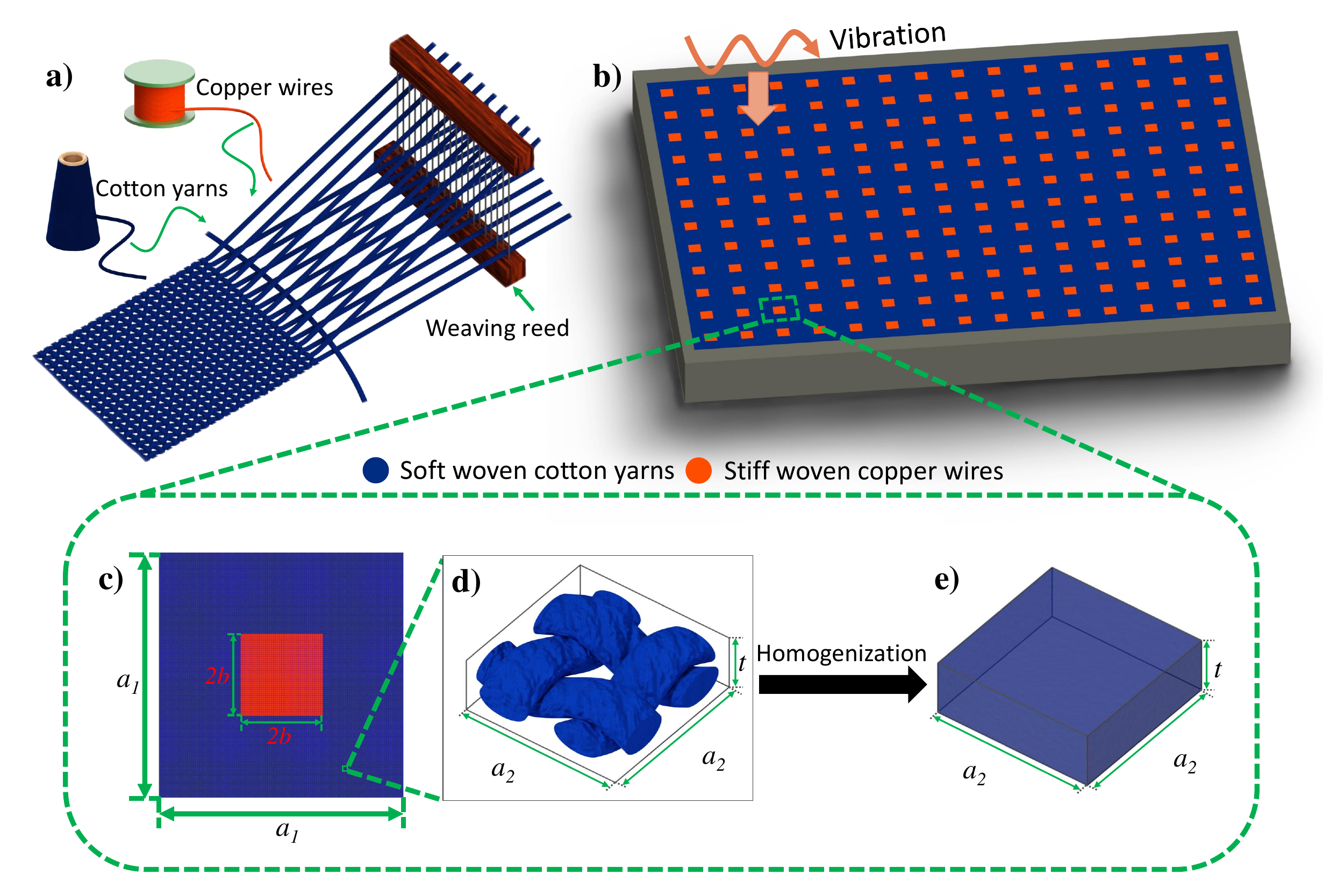}
\caption{\label{fig:Fig. 1} Overview of fabric phononic crystal design. a) A schematic of the weaving process used to fabricate the phononic crystal from cotton yarns and copper wires. (See Supporting Information for details.) b) A schematic of fabric phononic crystal for vibration control. c) A unit cell of the fabric phononic crystal, formed by woven copper surrounded by woven cotton. d) A single weave unit of plain woven cotton formed by warp and weft yarns, with dimensions of of $2.2~\text{mm} \times 2.2~\text{mm} \times 1.5~\text{mm}$, more than 30 times smaller than the macroscopic periods of the fabric phononic crystals in this work. e) Homogenized weave unit.}
\end{figure}

The schematic of a fully woven fabric phononic crystal is illustrated in Figures~\ref{fig:Fig. 1}. The macroscale unit cell consists of a square patch of woven copper wires at the center, surrounded by a woven cotton patch. In this study, the cotton fabric is selected as a soft material and the copper wire fabric as a stiff material. The key requirement is a large contrast in their mechanical properties, particularly density and Young’s modulus. 
Each cotton or copper patch comprises arrays of plain‑weave units whose characteristic dimensions are more than ten times smaller than the macroscale unit cell. Due to the broad range of length scales in these hierarchical fabric structures, fully resolved numerical modeling of a large sheet with yarn‑level detail is computationally prohibitive. To address this challenge, we developed a multi‑step homogenization procedure that sequentially captures the mechanical behavior at each structural scale. The homogenization procedure is a combination of experimental characterization with numerical modeling and dispersion relation fitting. 

Geometric models of plain‑weave units are generated by modeling the yarn cross‑sections as ellipses and sweeping the warp and weft yarn trajectories as sine curves \cite{gong2009modeling, turan2012variation}. Each weave unit has the dimensions of $2.2~\text{mm} \times 2.2~\text{mm} \times 1.5~\text{mm}$. As illustrated in Figure~\ref{fig:Fig. 1}c)–d), the objective of the homogenization procedure is to determine an equivalent set of mechanical properties for each cotton or copper weave unit. This enables each weave unit to be represented as an equivalent block with spatially uniform yet anisotropic material properties, an approximation that is valid when the elastic wavelengths of interest are much larger than the characteristic dimensions of the weave pattern \cite{carvelli2001homogenization, thierry2021homogenisation, thierry2018multi, thierry2020experimental}. Different homogenization strategies are adopted for the copper and cotton weave units, reflecting their different structural characteristics.

\subsubsection{Homogenization of the copper weave unit}

The copper patch is treated as a stiff, quasi‑rigid inclusion within the phononic crystal. Given the high stiffness of copper relative to cotton, the copper weave unit is homogenized using a simplified approach.
The effective mass density of the copper patch is calculated as the volume‑averaged density accounting for the volume fraction of copper wires relative to air within the unit cell. The effective Young’s modulus is assigned directly as the bulk Young’s modulus of copper, neglecting compliance induced by the weave geometry. This approximation is justified because the elastic deformation of the copper weave is negligible compared to that of the surrounding cotton fabric at low vibration frequencies (e.g., 1\textendash100 Hz in our study). We used two types of copper wires in our work, AWG 24 copper with a diameter of $0.51 \text{mm}$ and AWG 26 copper with a diameter of $0.41\text{mm}$. The effective properties of the corresponding copper weave units are listed in Table~\ref{table:copper}.

\renewcommand{\arraystretch}{1.3}
\begin{table}[h]
\centering
\caption{Effective properties of homogenized copper weave units}
\begin{tabular}{c  c  c  c  c  c  c}
\hline
\multirow{2}{*}{AWG 24 Copper weave unit} 
& $E$ (N/m$^2$) & $\nu$ & $G$ (N/m$^2$) & $\rho_{eff}$ (kg/m$^3$) & & \\ \cline{2-7}
& $110 \times 10^9$ & 0.35 & $4.074 \times 10^{10}$ & 7880 & & \\
\hline
\multirow{2}{*}{AWG 26 copper wave unit} 
& $E$ (N/m$^2$) & $\nu$ & $G$ (N/m$^2$) & $\rho_{eff}$ (kg/m$^3$) & & \\ \cline{2-7}
& $110 \times 10^9$ & 0.35 & $4.074 \times 10^{10}$ & 5254 & & \\
\hline
\end{tabular}
\label{table:copper}
\end{table}

\subsubsection{Homogenization of the cotton weave unit}


Homogenization of the cotton weave unit begins at the yarn scale. Individual cotton yarns are modeled as orthotropic beams, reflecting their fibrous microstructure.
The axial Young’s modulus of the cotton yarn ($E_1$) is determined experimentally as $E_1 = 1.42$~GPa (details in the Experimental section). The density of bulk cotton, which is $1440~\text{kg}/\text{m}^3$, is assigned as the the density of the cotton yarn. 

At the weave‑unit scale, the cotton fabric is modeled as a transversely isotropic material, reflecting its approximately symmetric in‑plane behavior and distinct out‑of‑plane response.
The in‑plane Young’s moduli ($E_1$ and $E_2$) and shear modulus ($G_{12}$) of the woven cotton fabric are experimentally measured as $E_1 = E_2 = 286.02~\text{MPa}$ and $G_{12} = 94.29~\text{MPa}$ (details in the Experimental section). The effective density of the homogenized block is calculated as the
volume‑averaged density.


Most existing homogenization studies of woven fabrics assume that the textile is embedded within a surrounding matrix material, which provides well‑defined boundary conditions and enables direct retrieval of effective properties \cite{carvelli2001homogenization, thierry2021homogenisation, thierry2018multi, thierry2020experimental}. In contrast, the fabric phononic crystal considered here contains no matrix material and exists as a free‑standing structure. The absence of a planar boundary layer makes it difficult to define an effective thickness and to directly measure the full set of three‑dimensional anisotropic elastic constants. Moreover, reliable experimental loading in the out‑of‑plane direction is not feasible for unsupported fabric samples. As a consequence, non‑axial and out‑of‑plane elastic constants cannot be obtained directly from standard mechanical tests. These remaining effective material parameters are therefore estimated by fitting the dispersion relations of the homogenized block model to those of the detailed, yarn‑resolved weave‑unit model. 

\begin{figure}[ht!]
\centering
\includegraphics[width=1\textwidth]{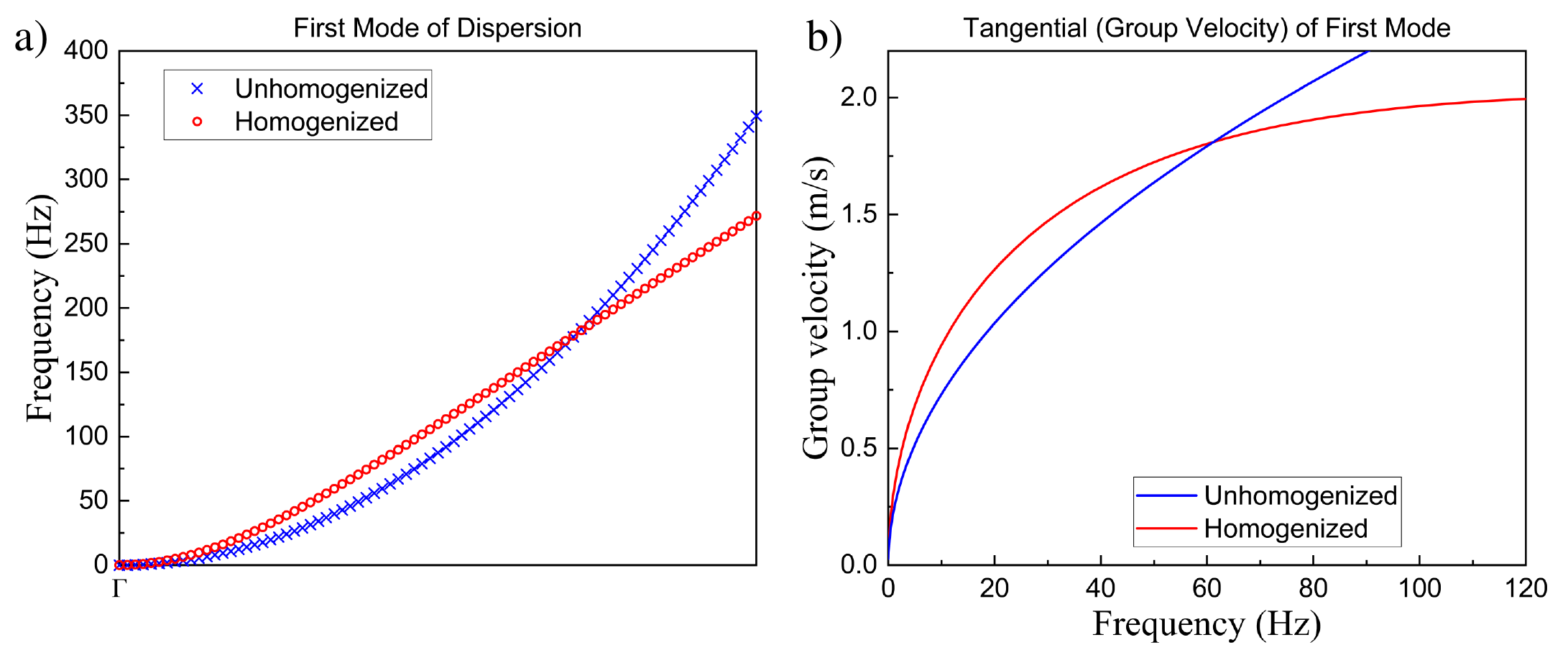}
\caption{\label{fig:Fig. 2} Comparisons of dispersion relations and wave group velocities for unhomogenized and homogenized cotton weave units. a) Dispersion relations of the unhomogenized cotton weave unit (blue cross) and the homogenized cotton block (red circle). Only the first branch of the out-of-plane vibration mode along 110 direction is shown. b) The tangential of the two dispersion branches in 1), representing the out-of-plane vibration wave group velocities for the unhomogenized cotton weave unit (blue) and the homogenized cotton block (red). The crossing of the two curves indicates the same group velocity.}
\end{figure}

The dispersion relations of both the unhomogenized (yarn‑level) weave unit and the homogenized block model are computed using COMSOL Multiphysics 6.1. Figure~\ref{fig:Fig. 2}a) compares the resulting band structures, where only the first branches of the out‑of‑plane modes are shown. The blue cross and red circles represent the dispersion relation of the unhomogenized and homogenized weave unit, respectively. Some discrepancies between the two dispersion relations are observed. These differences arise because complex yarn‑to‑yarn contact, local bending, and interlacing effects cannot be fully captured by the simplified homogenized block model. Similar discrepancies between unhomogenized and homogenized dispersion relations have been reported in prior multiscale studies of woven fabrics \cite{thierry2018multi, thierry2020experimental}. Rather than attempting to match the dispersion curves point‑by‑point over the entire frequency range, we adopt a physically motivated fitting strategy based on group velocity. Specifically, we compare the group velocities obtained from the slope of the first out‑of‑plane dispersion branch, and adjust the effective material parameters of the homogenized block to minimize discrepancies. In Figure~\ref{fig:Fig. 2}b, the blue and red curves represent the group velocities in the unhomogenized and homogenized models, respectively. The two group velocity curves intersect at a frequency of approximately 60~Hz, indicating that the two models exhibit identical wave‑propagation behavior at this frequency. This fitting criterion allows us to neglect higher‑frequency deviations while ensuring accurate representation of the dominant dynamic behavior relevant to the operating frequency range of interest.

The measured and fitted effective mechanical properties of the cotton yarns are summarized in Table~\ref{table:materialyarn}. The measured and fitted effective mechanical parameters of the homogenized cotton and copper blocks is summarized in Table~\ref{table:homogenized}. These effective properties in Table~\ref{table:homogenized} are subsequently used to model macroscopic phononic‑crystal unit cells without explicitly resolving yarn‑level geometry.

\renewcommand{\arraystretch}{1.3}

\begin{table}[h]
\centering
\caption{Effective properties of the cotton yarn}
\begin{tabular}{c  c  c  c  c  c  c}
\hline
$E_1$ (N/m$^2$) & $E_2$ (N/m$^2$) & $E_3$ (N/m$^2$) & $\nu_{12}$ & $\nu_{13}$ & $\nu_{23}$ \\ 
\hline
$1.42 \times 10^{9}$ & $9.5 \times 10^{7}$ & $9.5 \times 10^{7}$ & 0.35 & 0.35 & 0.2 \\
\hline
$G_{12}$ (N/m$^2$) & $G_{13}$ (N/m$^2$) & $G_{23}$ (N/m$^2$) & $\rho$ (kg/m$^3$) &   &  \\ 
\hline
$5.24 \times 10^8$ & $5.24 \times 10^8$ & $3.54 \times 10^7$ & 1440 &  &  \\ 
\hline
\end{tabular}
\label{table:materialyarn}
\end{table}

\begin{table}[h]
\centering
\caption{Effective properties of homogenized cotton weave units}
\begin{tabular}{c  c  c  c  c  c  c}
\hline
&$E_1$ (N/m$^2$) & $E_2$ (N/m$^2$) & $E_3$ (N/m$^2$) & $\nu_{12}$ & $\nu_{13}$ & $\nu_{23}$ \\ \cline{2-7}
&$2.86 \times 10^{8}$ & $2.86 \times 10^{8}$ & $2.86 \times 10^{5}$ & 0.41 & 0.1 & 0.1 \\ \cline{2-7}
&$G_{12}$ (N/m$^2$) & $G_{13}$ (N/m$^2$) & $G_{23}$ (N/m$^2$) & $\rho_{eff}$ (kg/m$^3$) & & \\ \cline{2-7}
&$9.4 \times 10^7$ & $9.4 \times 10^4$ & $9.4 \times 10^4$ & 600 &  & \\
\hline
\end{tabular}
\label{table:homogenized}
\end{table}

\subsection{Design of fabric phononic crystal as a bandstop filter}

\begin{figure}[ht!]
\centering
\includegraphics[width=1\textwidth]{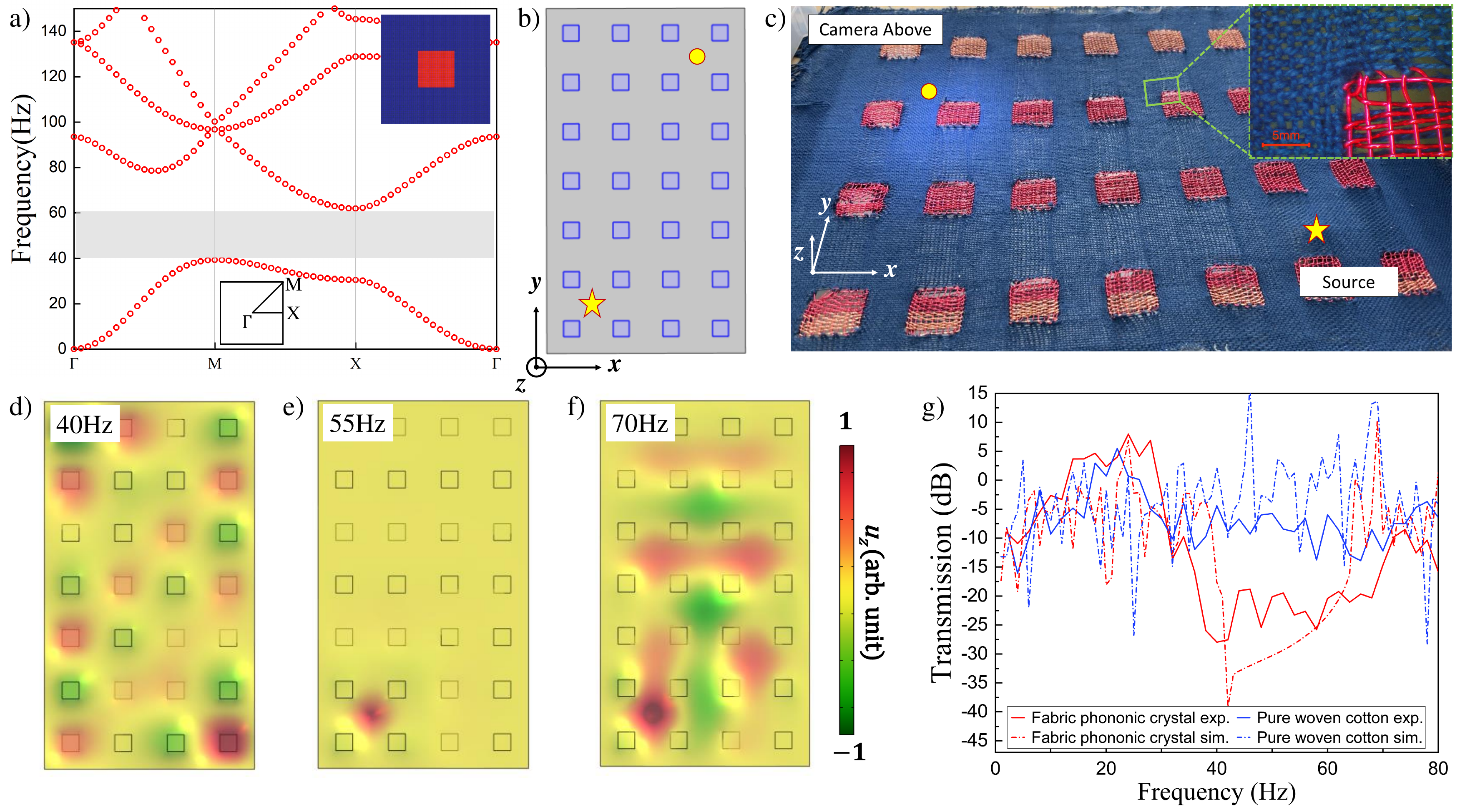}
\caption{\label{fig:Fig. 3} Fabric bandstop filter design and experimental characterization. a) Dispersion relation of the fabric phononic crystal unit cell for bandstop filtering. The gray shade represents the bandgap with a frequency range of $44.69$~Hz to $62.72$~Hz. b) Schematic of the fabric bandstop filter with $4 \times 7$ periods, each having dimensions of $76.2~\text{mm}\times76.2~\text{mm}\times1.5~\text{mm}$. The gray and blue regions represent the woven cotton and woven copper, respectively. The vibration source is placed at the lower left corner marked by the star, and the observation point is placed at the opposite corner marked by the circle. c) A picture of the fabricated fabric bandstop filter fixed to a hardboard frame. The shaker, which serves as the vibration source, is positioned beneath the fabric sheet at the star‑marked location. The high-speed camera for vibration observation and recording is located at the opposite corner marked by the circle. The inset figure shows the connection of woven cotton yarns and copper wires at the woven-cotton interface. d), e), and f) Simulated mode shapes at the frequencies of $40$~Hz (d), $55$~Hz (e), and $70$~Hz (f), respectively. g) Experimentally measured (solid) and simulated (dashed) vibration transmission curves of the fabric bandstop filter (red) and pure woven cotton sheet (blue) of the same size.}
\end{figure}

We first design and demonstrate a fabric bandstop filter for out-of-plane vibrations. The filter is based on a periodic pattern formed by woven cotton and woven copper patches. The cotton and copper regions are illustrated by the blue and orange areas in Figure~\ref{fig:Fig. 1}b), respectively. The macroscopic unit cell has dimensions of $76.2~\text{mm}\times76.2~\text{mm}\times1.5~\text{mm}$ ($a_1\times a_1\times t$). The copper patch at the center of the unit cell has dimensions of $25.4~\text{mm}\times25.4~\text{mm}\times1.5~\text{mm}$ ($2b\times2b\times t$). The dispersion relation of the unit cell,  as shown in Figure~\ref{fig:Fig. 3}a), were calculated using COMSOL 6.1 (details are provided in the Experimental Section). The shaded region indicates a bandgap from $44.69$~Hz to $62.72$~Hz. Within this frequency range, the vibration is localized near the excitation location and cannot propagate through the fabric.

We then simulated a finite-sized fabric phononic crystal with fixed boundaries. The fabric phononic crystal consists of $4\times7$ unit cells, as shown in Figure~\ref{fig:Fig. 3}b). A point source is placed at the lower-left corner. It generates out-of-plane vibrations using a frequency sweep from 1~Hz to 120~Hz in 1~Hz steps. A second point on the opposite side of the fabric phononic crystal is selected as the observation point. We calculated the transmission coefficient at each frequency. It is defined as the ratio of the vibration amplitude at the observation point to that at the source. The result is plotted as the red dashed curve in Figure~\ref{fig:Fig. 3}g). The transmission coefficient shows a clear bandgap from 43~Hz to 63~Hz. Within this range, the transmitted vibration amplitude decreases by $\sim35$ dB compared with the passbands. We further validated the results by examining the simulated vibration field distributions at three frequencies: 40~Hz, 55~Hz, and 70~Hz. These frequencies are before, inside, and after the bandgap in the transmission curve (Figure~\ref{fig:Fig. 3}d)–f)). At 55~Hz, very little vibration propagates through the fabric phononic crystal, as shown in Figure~\ref{fig:Fig. 3}e). The vibration remains localized near the source. In contrast, at 40~Hz and 70~Hz, the vibration propagates across the crystal, as shown in Figure~\ref{fig:Fig. 3}d) and Figure~\ref{fig:Fig. 3}f), respectively.

We fabricated the designed fabric phononic crystal using a double-weaving technique. This method weaves 10/2 cotton yarns and AWG 24 copper wires at the same time on a weaving loom. It produces a fully woven rectangular fabric phononic crystal sheet with $4\times7$ unit cells, as shown in Figure~\ref{fig:Fig. 3}c). Details of the fabrication are provided in the Supporting Information.

To further illustrate the bandgaps, we experimentally measured vibration transmission across the fabric phononic crystal. Details of the measurements are provided in the Experimental Section. The measured transmission coefficient of the fabric phononic crystal (red solid curve in Figure~\ref{fig:Fig. 3}g)) shows a strong drop in vibration transmission from 45 Hz to 65 Hz. This result indicates the existence of a bandgap. This frequency range also matches well with the simulated bandgap location.

For comparison, we also simulated and experimentally measured the transmission coefficient of a pure cotton woven patch with the same size as the fabric phononic crystal. The results are plotted as the blue dashed (simulation) and blue solid (experiment) curves in Figure~\ref{fig:Fig. 3}g). No obvious amplitude decrease appears in the same bandgap range. This indicates that the bandgap exists only in the fabric phononic crystal. The bandstop filtering behavior is attributed to the periodic pattern of woven cotton and woven copper in the fabric phononic crystal structure.

\subsection{Design of fabric phononic crystal as a topological insulator}

\begin{figure}[ht!]
\centering
\includegraphics[width=0.8\textwidth]{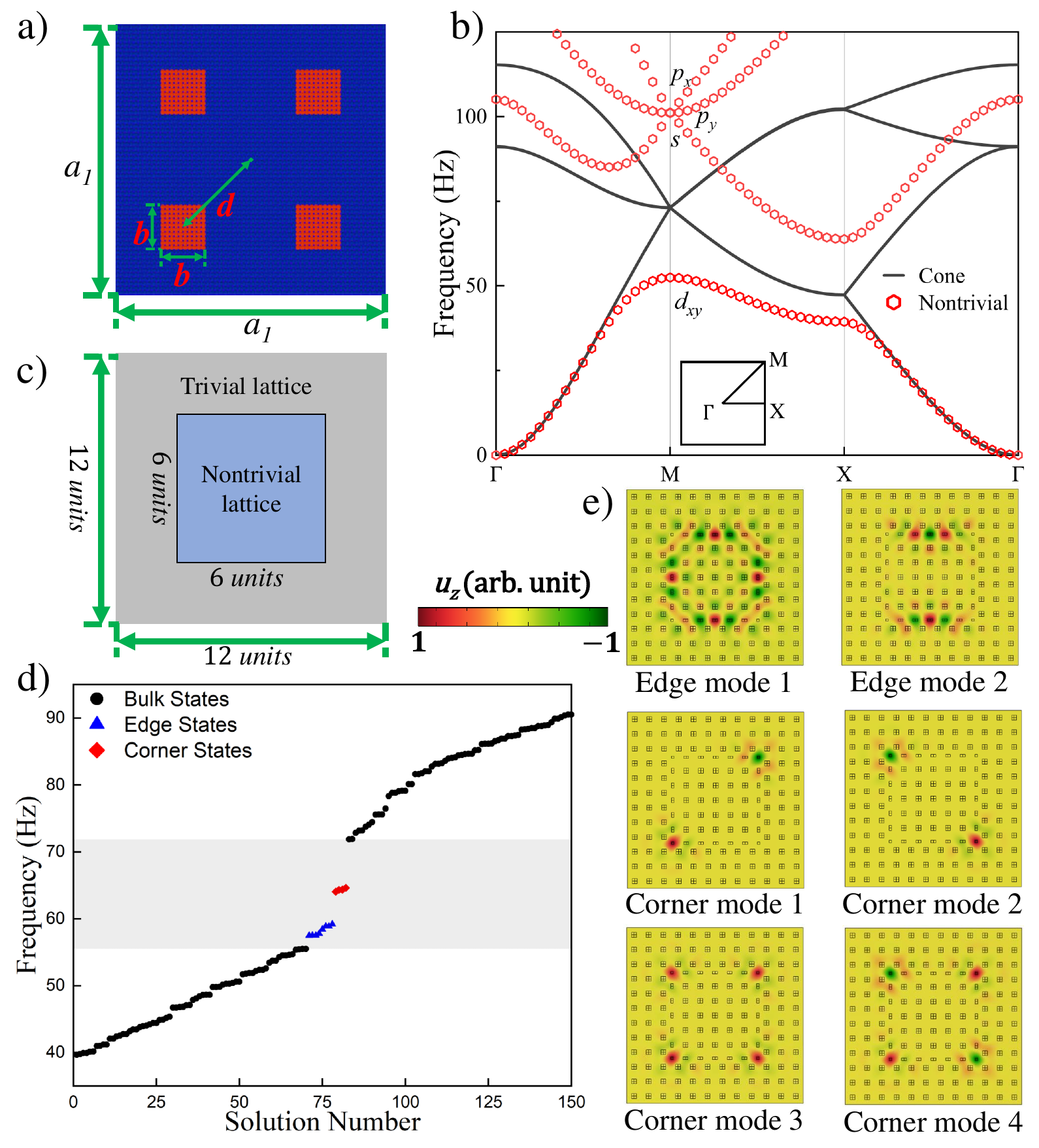}
\caption{\label{fig:Fig. 4} Design of the fabric topological insulator. a) A unit cell of the fabric topological insulator with the dimensions of $76.2~\text{mm}\times76.2~\text{mm}\times1.5~\text{mm}$. There are four woven copper patches inside a single unit cell with the dimensions of $12.7~\text{mm}\times12.7~\text{mm}\times1.5~\text{mm}$. b) Dispersion relations of lattices. The solid black line represents the dispersion relation when $d=\sqrt2 a_1/4$, which is the geometry in a), showing a cone-like band structure. The red hollow hexagon represents the dispersion relation of a trivial lattice when $d=5\sqrt2 a_1/12$, showing a bandgap. c) A schematic of the designed topological insulator, including $12 \times 12$ units in total with a nontrivial lattice with $6 \times 6$ units at the center, surrounded by a trivial lattice. d) The eigenfrequency spectrum of the phononic crystal structure shown in c), where the bulk, edge, and corner states are plotted as black circles, blue triangles, and red diamonds, respectively. e) The mode shapes of edge and corner states. The first row shows two edge states, with vibrational energy localized either along all four edges or along a pair of opposite edges. The second row presents two corner states, in which vibrations are localized at diagonally opposite corners with opposite phases. The third row shows corner states with vibrational localization at all four corners.}
\end{figure}

Building upon the phononic bandgap established in the designed fabric structure, we further investigate its capability to support topological edge and corner states beyond conventional bandgap formation through modification of the unit‑cell geometry \cite{zheng2023switchable}. The redesigned unit cell, shown in Figure~\ref{fig:Fig. 4}a), consists of woven copper inclusions embedded within a woven cotton matrix. The unit‑cell size is kept identical to that of the structure examined in the previous section. Four copper inclusions were incorporated into each unit cell, each with a side length equal to one‑half that of the earlier design. By systematically varying the offset distance between the copper square inclusions and the center of the unit cell (the $d$ in Figure~\ref{fig:Fig. 4}a), spatial inversion symmetry of the lattice is either broken or restored. This controlled symmetry tuning enables transitions between distinct topological phases, thereby allowing the emergence or suppression of topologically protected edge and corner states.

When $d=\sqrt2 a_1/4$, the resulting band structure is plotted in Figure~\ref{fig:Fig. 4}b) as solid black lines, revealing a Dirac cone degeneracy at the $M$ point. A bandgap appears when $d$ deviates from $d=\sqrt2 a_1/4$, as illustrated by the red-dotted dispersion relation in Figure~\ref{fig:Fig. 4}b) when $d=5\sqrt2 a_1/12$. When $d<\sqrt2 a_1/4$, it is a trivial lattice, and when $d>\sqrt2 a_1/4$, it is a non-trivial lattice. In addition, when $d$ is set to the value of $\sqrt2 a_1/12$ or $5\sqrt2 a_1/12$, the copper squares with a side length of $b$ connect to each other at the center or corner of the unit cell, respectively, forming a square with a side length of $2b$. This geometry corresponds to the geometry of the fabric phononic crystal in the previous section in Figure~\ref{fig:Fig. 1}c). 

To investigate higher‑order topological phenomena in the fabric phononic crystal, we analyzed the band inversion of the woven lattice using an effective Hamiltonian description (See Supporting Information for details). The theory shows that tuning the inclusion geometry inverts the $s$ and $d_{xy}$ bands at the M point, which changes the bulk Dirac mass and marks the transition between trivial and nontrivial phases. In a finite woven phononic crystal, this bulk inversion gives rise to in-gap edge states and localized corner modes. To show this, we analyzed the eigenfrequency spectrum of a finite phononic crystal as shown in Figure~\ref{fig:Fig. 4}c). It consists of $12 \times 12$ unit cells. A central $6 \times 6$ nontrivial lattice is surrounded by a trivial lattice. The resulting eigenfrequency spectrum is shown in Figure~\ref{fig:Fig. 4}d), where the bulk, edge, and corner states are identified. The corresponding mode shapes are shown in Figure~\ref{fig:Fig. 4}e). They confirm the presence of both one-dimensional edge states and zero-dimensional corner states. These features are characteristic signatures of second-order topological insulators \cite{zheng2023switchable, zhang2019second}. 

\begin{figure}[ht!]
\centering
\includegraphics[width=1\textwidth]{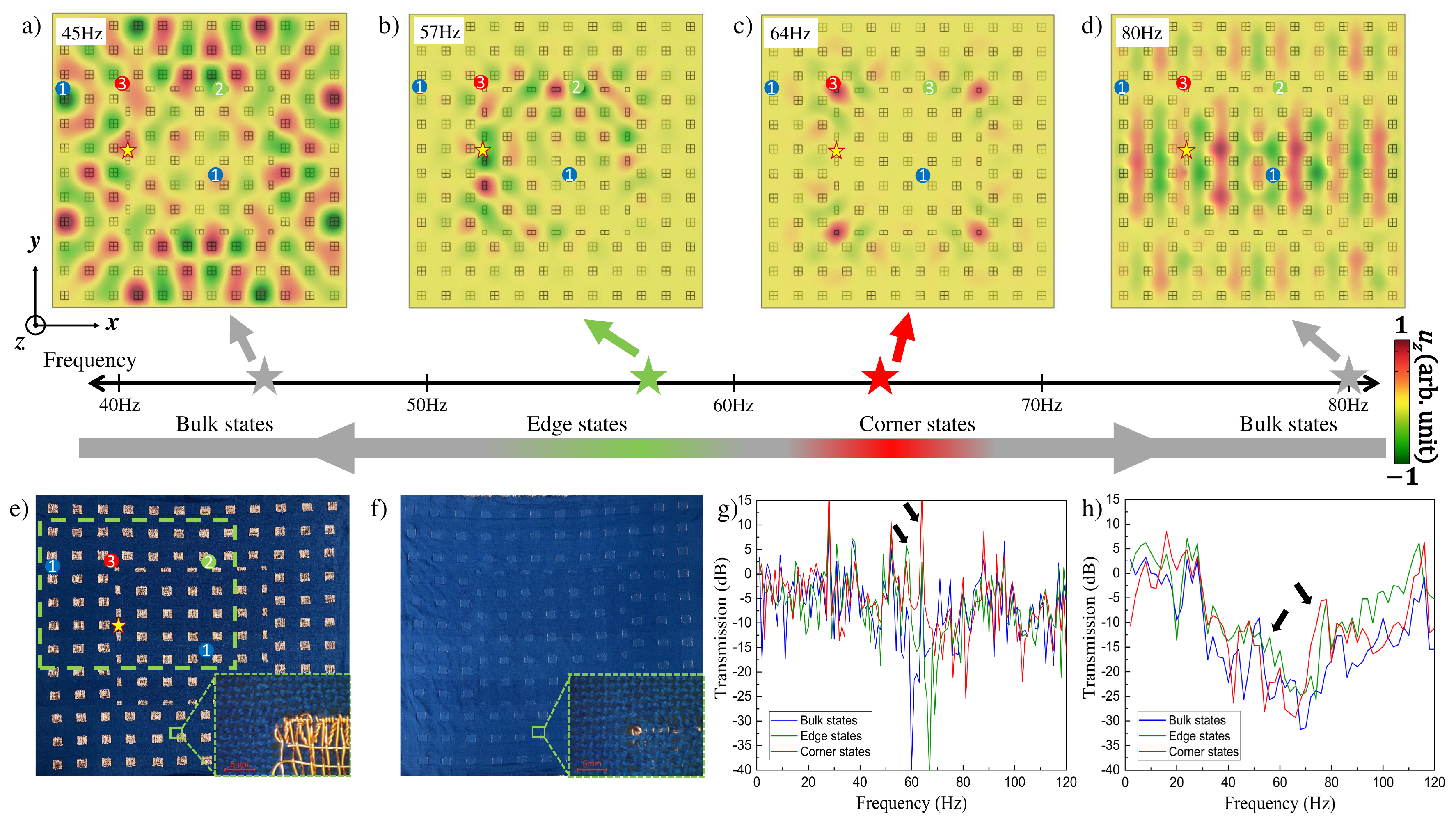}
\caption{\label{fig:Fig. 5} Vibration propagation simulation and experimental measurements for the fabric topological insulator. a)\textendash d) Simulated mode shapes of the bulk state, topological edge state, topological corner state, and another bulk state at frequencies of  $45$~Hz, $57$~Hz, $64$~Hz, and $80$~Hz, respectively. e), f) Images of the front (e) and back (f) of the fabricated fabric topological insulator sheet, fixed onto a hardboard frame. The vibration source is located beneath the star-marked location. The labels ``1--3'' represent the measurement locations for the bulk states, topological edge state, and the topological corner state, respectively, same as the source and measurement locations in simulation. Insets show the details of the front and back inside a single unit cell. g), h) The simulated and experimentally measured transmission spectra at four measurement locations, respectively. Arrows are pointed at the locations of the edge and corner states.}
\end{figure}

Furthermore, we conducted vibration propagation simulations for the same fabric phononic crystal as in Figure~\ref{fig:Fig. 4}c), with an external vibration source. The simulated vibration field distributions are shown in Figure~\ref{fig:Fig. 5}a–d). The source location is marked by a star.  We probed vibrations in the bulk states of trivial and nontrivial lattices, edge state, and corner state, respectively. The resulting transmission spectra are shown in Figure~\ref{fig:Fig. 5}g). The relatively low transmission from 55~Hz to 80~Hz corresponds to the bulk bandgap in both the trivial and nontrivial lattices. The relatively high transmission at $57$~Hz indicates the edge state. Its vibration distribution is shown in Figure~\ref{fig:Fig. 5}f). The peak near $64$~Hz corresponds to the corner state, as confirmed by the vibration distribution in Figure~\ref{fig:Fig. 5}c). In addition, at $45$~Hz and $80$~Hz, which fall in the bulk passbands of the trivial and nontrivial lattices, the out-of-plane vibration generated by the source propagates throughout the entire fabric phononic crystal, as shown in Figure~\ref{fig:Fig. 5}a) and Figure~\ref{fig:Fig. 5}d).

We then fabricated the topological insulator using the same double-weaving technology with 10/2 cotton yarns and AWG 26 copper wires. The sample follows the design shown in Figure~\ref{fig:Fig. 4}. The fabricated sample is shown in Figure~\ref{fig:Fig. 5}e) and f). It contains $12 \times 12 = 144$ unit cells in total. The central $6 \times 6 = 36$ unit cells form the nontrivial lattice, and the surrounding unit cells form the trivial lattice.

The vibration characterization of the fabricated fabric topological insulator was also performed. The phononic crystal was fixed to a hardboard frame and placed horizontally. We used the same vibration source and measurement locations as in the simulations, as shown in Figure~\ref{fig:Fig. 5}e). A high-speed camera was placed above the source location and the four observation locations in sequence to measure the vibration. The input signal settings and the signal processing procedure were the same as those used to characterize the bandstop filter in the previous section. The resulting transmission spectra are shown in Figure~\ref{fig:Fig. 5}h). The relatively high transmission near $60$~Hz indicates the presence of the topological edge state. The peak near $72$~Hz corresponds to the topological corner state. Meanwhile, the low transmission from $50$~Hz to $80$~Hz represents the bulk bandgap in both the trivial and nontrivial lattices. By comparing the simulated and experimental transmission spectra in Figure~\ref{fig:Fig. 5}g) and h), we observe that the experimental amplitudes differ from the simulation and show a slight frequency shift. This difference is attributed to the effects of damping, gravity, and fabric tension during measurement, as well as slight nonuniformities in the distribution of the warp and weft cotton yarns and copper wires. 

\section{Conclusion}

In this work, we established fully woven fabrics as a practical and scalable platform for phononic crystals that can passively filter and control elastic vibrations. By integrating a soft cotton weave with stiff woven copper inclusions, we created hierarchical fabric architectures whose periodicity and material contrast produce nontrivial dispersion behavior. An important enabler is a multiscale computational framework for the predictive design of woven systems. It combines homogenized mechanical properties computed at the weave scale with a macroscale model that applies these properties to simulate large fabric architectures without explicitly resolving yarn‑level details. This approach reduces the computational cost of full-resolution modeling and enables efficient analysis of periodic fabric lattices. The current framework also has limitations. It does not fully capture yarn contacts, friction, and sliding, and simplifies local yarn deformation and rearrangement under loading. These effects may contribute to differences between simulations and experiments, including amplitude mismatch and frequency shifts.

Building on this platform, we demonstrate two types of phononic functionality. First, we realize and validate phononic bandgaps in fully woven fabric phononic crystals, confirmed through both multiscale simulations and vibration-transmission experiments. Second, we extend this concept to topological wave control by designing a woven architecture that supports higher-order topological states, including in-gap edge modes and localized corner modes. The agreement between modal predictions, transmission spectra, and measured vibration fields verifies that phononic functionalities can be encoded directly into the fabric geometry and material layout. 

More broadly, this work expands the scope of phononic-crystal-based wave manipulation into fully woven textile materials. It identifies fabrics as an emerging class of metamaterial systems in which wave control and vibration signal processing arise from material architecture rather than dense arrays of active components. The demonstrated concepts open opportunities for passive and conformable vibration-management layers, acoustic and vibration sensing elements, wearable sensing and haptic interfaces, noise‑mitigation textiles, and other engineered fabric systems that integrate wave control functionalities within a single woven form factor.

\section{Experimental Section}

\subsection{Dispersion relation calculations}

The dispersion relations were calculated in COMSOL Multiphysics using the Solid Mechanics module and eigenfrequency study. The fabric thickness used in each model was taken as the average of measurements obtained at multiple locations on the corresponding sample. For all periodic calculations, Floquet-Bloch periodic boundary conditions were imposed on opposite boundaries of the unit cell, and the Bloch wave vector was swept along the boundary of the first Brillouin zone to construct the dispersion relations, following the high-symmetry path $\Gamma - M - X - \Gamma$. The resulting eigenfrequencies were used to construct the dispersion relation. The mesh convergence was verified by progressively refining the mesh until the eigenfrequencies of the bands of interest changed negligibly. In the present study only out-of-plane vibration modes were considered, and the relevant modes were identified using the polarization-index method \cite{chaunsali2018experimental}.

The dispersion relation of the cotton weave unit, shown by the blue crosses in Figure~\ref{fig:Fig. 2}a), was calculated using the material properties of the yarn employed in the simulation, as listed in Table~\ref{table:materialyarn}. The dispersion relation of the homogenized cotton block, shown by the red circles in Figure~\ref{fig:Fig. 2}a), was obtained using the homogenized effective material properties of the cotton weave unit given in Table~\ref{table:homogenized}. In Figure~\ref{fig:Fig. 2}a), only the first out-of-plane band near the $\Gamma$ point is shown.
The dispersion relations of the macroscopic cotton–copper unit cell, shown in Figure~\ref{fig:Fig. 3}a) and Figure~\ref{fig:Fig. 4}b), were calculated using the homogenized effective material properties of the cotton and copper weave units listed in Table~\ref{table:homogenized}. The connection between woven cotton and copper was modeled as a direct rigid connection.

\subsection{Experimental characterization of cotton yarns and cotton fabric samples}

\begin{figure}[ht!]
\centering
\includegraphics[width=1\textwidth]{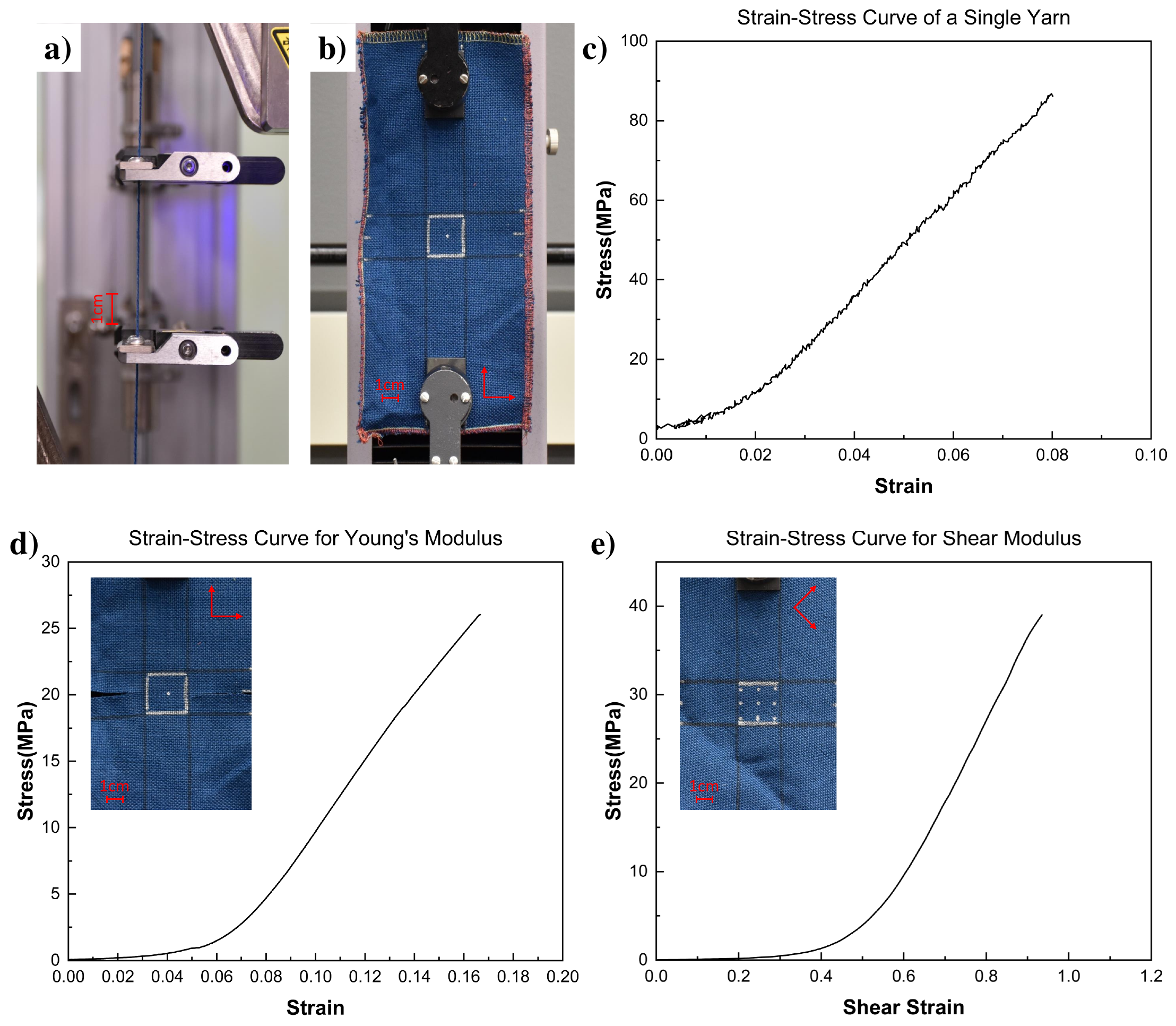}
\caption{\label{fig:Fig. 6} Experimental characterization of cotton yarns and cotton fabric samples. a) and c) Experimental setup for measuring the axial Young's modulus $E_1$ of a single 10/2 cotton yarn (a) and the corresponding averaged stress\textendash strain curve from repeated measurements (c). b) Experimental setup for measuring the in-plane Young's moduli $E_1$ and $E_2$ of a cotton fabric sample at a $0^\circ$ rotation, and the shear modulus $ G_{12}$ at a $45^\circ$ rotation. The measured fabric samples are in the dimensions of $100~\text{mm}\times200~\text{mm}$, and the clamps are at the center with a width of $25.4~\text{mm}$. d) and e) Averaged stress\textendash strain curves of the cotton fabric sample from multiple measurements. The weave directions of the fabric samples are indicated bu red arrows in b), d) and e).}
\end{figure}

Uniaxial tensile tests are performed on a single 10/2 cotton yarn using an Instron MTS C42 load frame equipped with a long‑travel XL extensometer, as shown in Figure~\ref{fig:Fig. 6}a). Five samples are tested to ensure repeatability. The averaged stress–strain response is shown in Figure~\ref{fig:Fig. 6}c), and the axial Young’s modulus is extracted from the linear elastic region of the curve as $ E_1 = 1.42$ GPa (standard deviation: $0.09 ~\text{GPa}$). %

The in‑plane Young’s moduli ($E_1$ and $E_2$) and shear modulus ($G_{12}$) of the woven cotton fabric are experimentally characterized according to ASTM D5034-21 using an Instron MTS C42 load frame equipped with a 5 kN load cell, as illustrated in Figure~\ref{fig:Fig. 6}b). Rectangular fabric samples with dimensions of $100~\text{mm} \times 200~\text{mm}$ are tested, with a clamp width of $25.4~\text{mm}$ centered on the sample. To extract directional properties, samples are oriented at $0^\circ$ and $45^\circ$ relative to the loading direction to evaluate $E_1$, $E_2$, and $G_{12}$, respectively. Three strain measurements are performed for each orientation. The averaged strain–stress curves are obtained by optical tracking the deformation of the surface reference lines or the angular changes of the central dot markers, as shown in Figure~\ref{fig:Fig. 6}d)–e). From the linear elastic regions of the averaged curves, the following properties are obtained:
$E_1 = E_2 = 286.02$ MPa (standard deviation: $19.31~\text{MPa}$), and $G_{12} = 94.29$ MPa (standard derivation: $7.23 ~\text{MPa}$).

\subsection{Measurement of the transmission spectra of fabric phononic crystals}

\begin{figure}[ht!]
\centering
\includegraphics[width=0.65\textwidth]{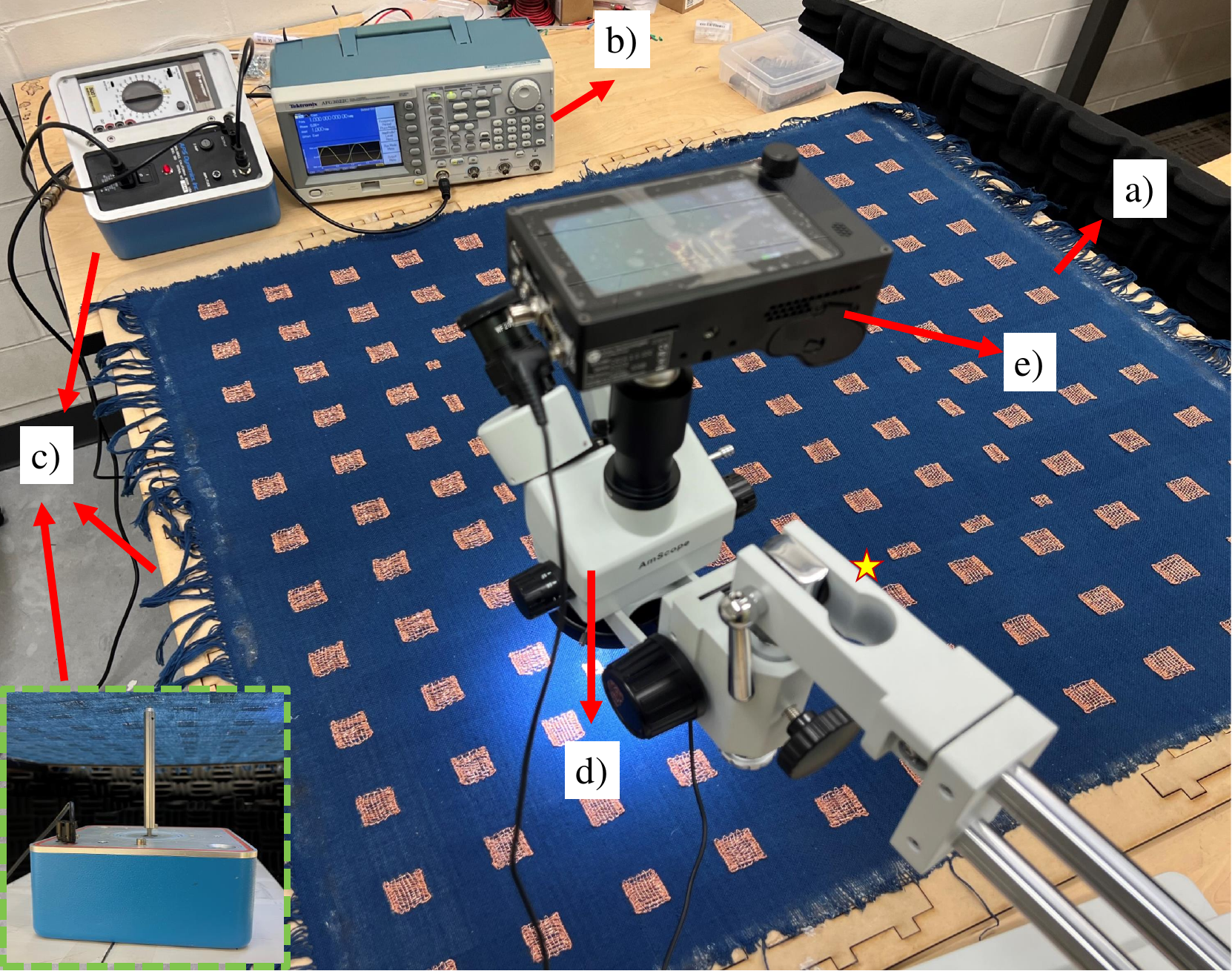}
\caption{\label{fig:Fig. 7} An photo of the experimental system for vibration transmission spectrum measurement. a) Horizontally placed fabric phononic crystal sheet on a rigid hardboard frame. b) Tektronix AFG2022C Function Generator. c) APS Dynamics Model 300 Shaker. The vibration source, shown in the inset, is placed under the phononic crystal at the star-marked location. d) AmScope optical microscope. e) Chronos 1.4 high-speed camera.}
\end{figure}

The fabricated fabric phononic crystal sheet is fixed to a rigid hardboard frame and placed horizontally, as shown in Figure~\ref{fig:Fig. 7}a). During mounting, small tension is applied along both in-plane directions to keep the sheet flat, reduce curvature due to gravity, and better preserve the designed geometry. A Tektronix AFG2022C function generator (Figure~\ref{fig:Fig. 7}b)) and an APS Dynamics Model 300 shaker (Figure~\ref{fig:Fig. 7}c)) are used as the vibration source. The shaker is oriented perpendicular to the fabric and excites the sample at the star-marked location. The input signal consists of continuous sinusoids at multiple center frequencies, $f_c$, spaced by 2 Hz from 2 Hz to 120 Hz. Vibration transmission is measured at the opposite corner of the phononic crystal using an AmScope optical microscope (Figure~\ref{fig:Fig. 7}d)) and a Chronos 1.4 high-speed camera (Figure~\ref{fig:Fig. 7}e)), consistent with the numerical setup. For each center frequency, 1000 frames are recorded at 1767 frames per second (fps). The pixel intensity at the observation location, extracted from the video frame sequence after grayscale conversion in MATLAB using the \texttt{rgb2gray} function, is used as the vibration signal and converted to a frequency spectrum. To reduce noise, 50 measurements at the same observation locations are performed at each $f_c$ and averaged. The vibration at the source location is also recorded and processed using the same procedure. The transmission coefficient spectrum is then calculated as the ratio of the spectral amplitudes at the observation point to those at the source for each $f_c$. The final transmission spectrum is obtained by averaging the transmission coefficient spectra over all center frequencies. The resulting transmission coefficients as functions of frequency are plotted in Figure~\ref{fig:Fig. 3}g) and Figure~\ref{fig:Fig. 5}h) for the bandstop filter and topological insulator, respectively.

\medskip


\section{Supporting Information}

\subsection{Details of the fabrication procedure for fabric phononic crystals}

\begin{figure}[ht!]
\centering
\includegraphics[width=1\textwidth]{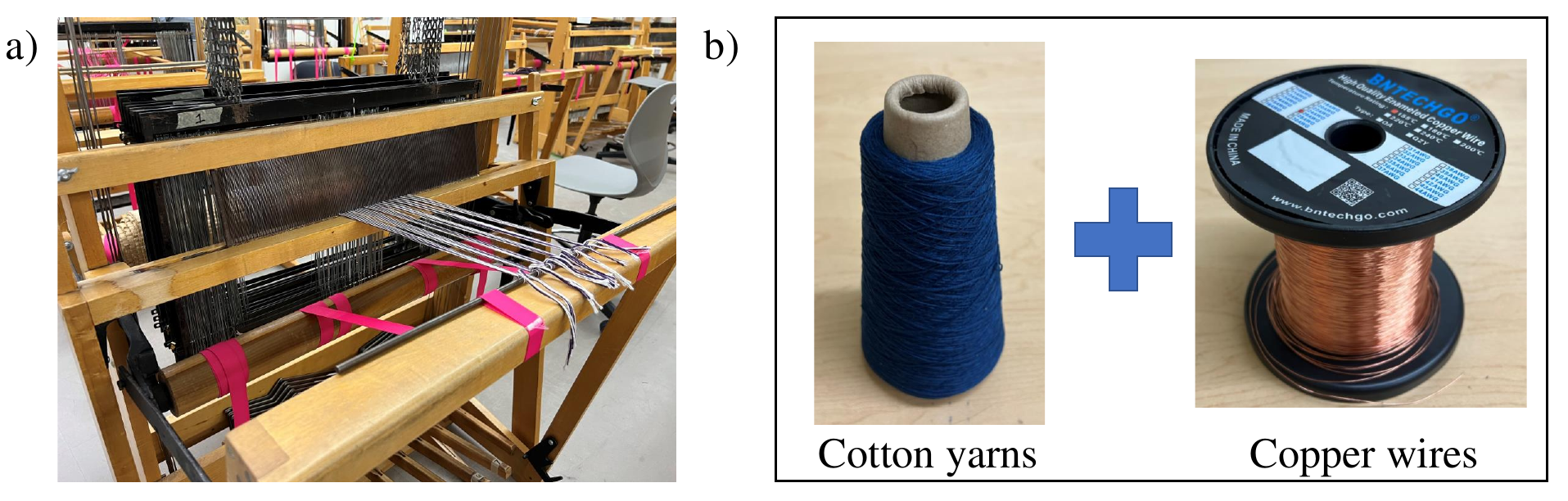}
\caption{\label{fig:Fig. S1} Fabrication of fabric phononic crystals. a) The weaving loom used for fabricating the fabric phononic crystals. b) Cotton yarns and copper wires used in fabric phononic crystals.}
\end{figure}

The fabric phononic crystals were constructed as two-dimensional fabric sheets in which square woven-copper inclusions were periodically embedded within a surrounding woven-cotton matrix. At the macroscale, each unit cell consisted of a square woven copper patch (or multiple woven copper patches) embedded within a square cotton patch. At the mesoscale (yarn level), both the cotton and copper regions were formed from arrays of plain-weave units. A representative cotton weave unit has the dimensions of $2.2~\text{mm} \times 2.2~\text{mm} \times 1.5~\text{mm}$. The bandstop-filter sample consists of $4 \times 7$ unit cells, while the fabric topological insulator consisted of $12 \times 12$ unit cells with a central $6 \times 6$ nontrivial lattice surrounded by a trivial lattice. 

The fabric phononic crystals were fabricated using a double-weaving process on a weaving loom. Photographs of the weaving loom and the cotton and copper wires are shown in Figure~\ref{fig:Fig. S1}. During weaving, the cotton warp yarns beneath the woven-copper regions were left at very low tension so that the woven copper patches remained mechanically attached between adjacent cotton patches after fabrication. Because these low-tension cotton yarns have much smaller mass than the copper wires, their contribution to out-of-plane vibration at low frequencies (below 100 Hz as in our study) was negligible and therefore omitted from the vibration analysis. After weaving, only the intended square woven-copper patches at their prescribed locations were retained, and all excess copper wires were manually removed. The same mesoscale geometry and weaving strategy were used to fabricate samples for Young’s modulus and shear modulus characterization.

Tension applied to both warp and weft yarns during weaving caused slight relaxation and shrinkage upon removal from the loom, yielding final dimensions marginally smaller than the nominal design. Minor irregularities were also introduced by nonuniform yarn and wire distributions and by the finite connection connection regions between the cotton and copper phases, which break the ideal symmetry of the pure-cotton and pure-copper regions. These fabrication-induced imperfections were small relative to the overall crystal dimensions and were therefore treated as structural defects with negligible influence on vibration behavior. They are not examined further. After fabrication and post-processing, the fabric phononic crystals were mounted on a rigid hardboard frame for experimental characterization.

\subsection{Band inversion when varying the distance of woven copper inclusions to the center of the unit cell}

\begin{figure}[ht!]
\centering
\includegraphics[width=0.8\textwidth]{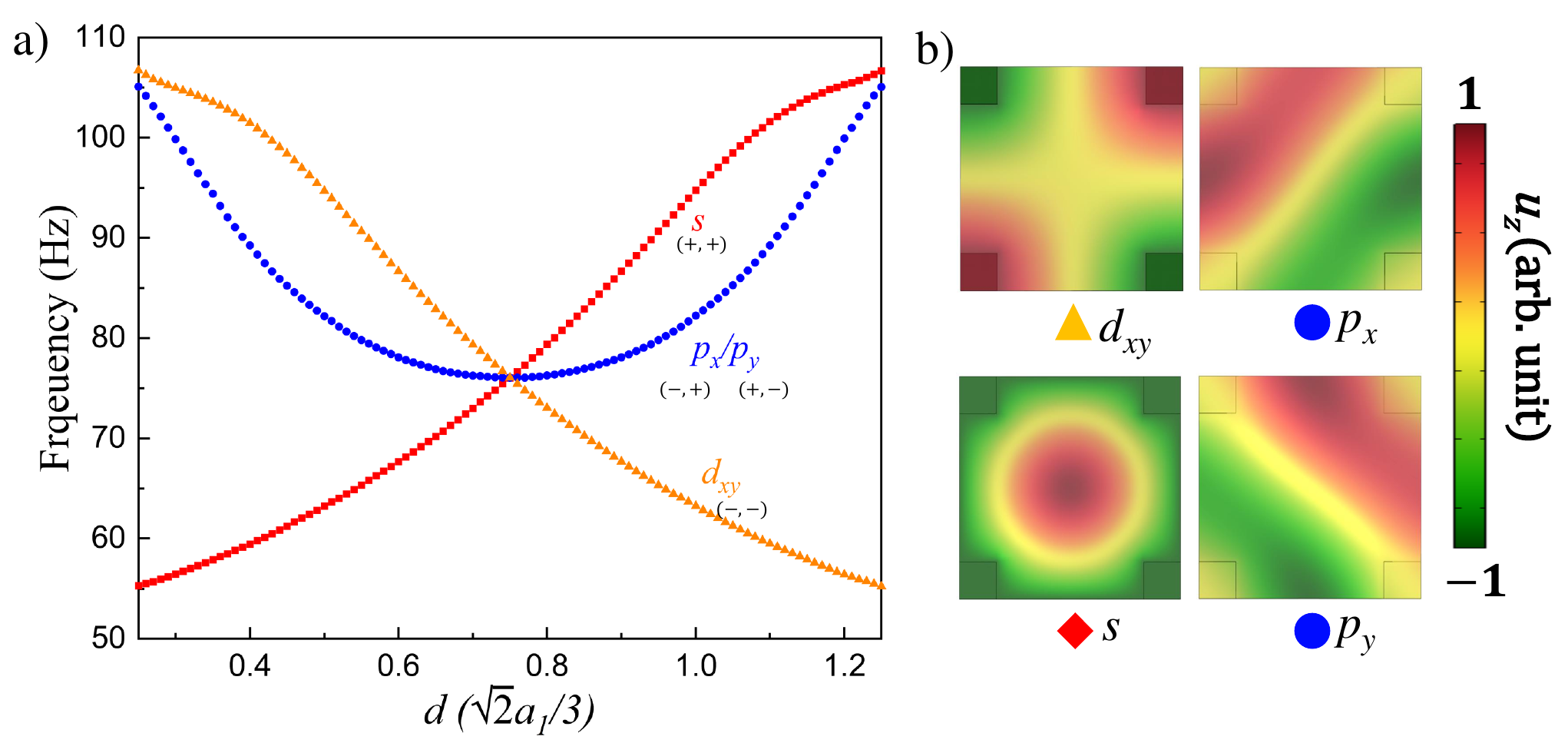}
\caption{\label{fig:Fig. S2} Illustration of dispersion band inversion when varying the distance $d$ of woven copper inclusions to the center of the unit cell. a) Plot of eigenfrequencies at ``M'' point in first Brillouin zone, showing band inversion of the $s$-like and $d$-like vibrational modes when altering $d$. The geometric parameters of $d = \sqrt2 a_1/12$ and $d = 5\sqrt2 a_1/12$ are used in simulation, fabrication and experiments. Different parity order is listed. b) Four eigenmodes of the corresponding bands for $d = 5\sqrt2 a_1/12$.}
\end{figure}

The eigenfrequencies at the ``M'' point in the first Brillouin zone (BZ) of the first four dispersion bands are shown in Figure ~\ref{fig:Fig. S2}a), showing the band inversion of s-like and d-like vibrational modes when varying the distance $d$ of woven copper inclusions to the center of the unit cell. The four eigenmodes of the corresponding bands with $d = 5\sqrt2 a_1/12$ are also shown in Figure ~\ref{fig:Fig. S2}b). The three fold degeneracy is observed at both $d=\sqrt2 a_1/12$ and $d = 5\sqrt2 a_1/12$. The band inversion between the $s$ and $d_{xy}$ modes originates from the variation of the effective stiffness and mass distribution in the unit cell. As the geometric parameter is tuned, the restoring force associated with the $s$-like mode and the nodal structure of the $d_{xy}$ mode respond differently, leading to opposite trends in their eigenfrequencies. Consequently, the two branches approach each other and eventually cross at the ``M'' point, resulting in an exchange of modal ordering. In addition, this inversion gives rise to the observed second-order topological states. The different parity order of each band is also marked in Figure ~\ref{fig:Fig. S2}a). Further details on the origin and implications of the band inversion are provided in the full Hamiltonian analysis.

\subsection{Hamiltonian theory for the fabric phononic topological insulators}

The transition of trivial and nontrivial lattices are driven by the displacements of the four woven copper inclusions to the unit cell center. This section aims to utilize the $\mathbf k \cdot \mathbf P$ theory \cite{kittel2018introduction} to understand the dispersions around the high symmetrical Point $M$.

The out-of-plane modes of the periodic structure satisfy a generalized eigenvalue equation
\begin{equation}
\hat H \psi_{n,\mathbf k}(\mathbf r)
=
\omega_{n,\mathbf k}^{\,2}\,
\hat W \psi_{n,\mathbf k}(\mathbf r)
\label{eq:S1}
\end{equation}
where $\psi_{n,\mathbf k}(\mathbf r)$ is the Bloch eigenmode, $\omega_{n,\mathbf k}$ is the eigenfrequency, $\hat H$ is the Hermitian operator of the system, and $\hat W$ is the corresponding weight operator. The Bloch eigenstates satisfy the orthogonality relation
\begin{equation}
\delta_{nn'}
=
\int_{\mathrm{u.c.}}
\psi_{n,\mathbf k}^{*}(\mathbf r)\,
\hat W\,
\psi_{n',\mathbf k}(\mathbf r)\,
d\mathbf r
\label{eq:S2}
\end{equation}
where u.c. denotes the unit cell, which is the integral over the entire unit cell.

Firstly, to study the dispersion near the degenerate bands, we perform a $\mathbf k \cdot \mathbf P$ expansion  around the high-symmetry point M in the first BZ, located at $\mathbf K=\left(\frac{\pi}{a},\frac{\pi}{a}\right)$. A small wavevector deviation is defined as $\mathbf q=\mathbf k-\mathbf K$, where $\mathbf{k}$ is the Bloch wavevector. This formulation captures the band behaviors near the $M$ point.

In the $\mathbf k \cdot \mathbf P$ theory, the Bloch eigenstate is expanded as
\begin{equation}
\psi_{n,\mathbf k}(\mathbf r)
=
e^{i\mathbf q\cdot \mathbf r}
\sum_{n'} C_{nn'}\,
\psi_{n',\mathbf K}(\mathbf r)
\label{eq:S4}
\end{equation}
Substituting Eq.~\ref{eq:S4} into Eq.~\ref{eq:S1} and expanding in powers of $\mathbf q$, the effective Hamiltonian can be written as
\begin{equation}
H(\mathbf q)=H^{(0)}+H^{(1)}+H^{(2)}+O(\mathbf q^3)
\label{eq:S5}
\end{equation}
where $H^{(0)}$, $H^{(1)}$, and $H^{(2)}$ denote the zeroth-, first-, and second-order terms, respectively. These terms in the effective Hamiltonian can be expaned as 
\begin{equation}
H_{nn'}(\mathbf q)
=
\omega_n^2 \delta_{nn'}
+
\mathbf q \cdot \mathbf P_{nn'}
+
\frac{1}{2}\,\mathbf q^{\mathrm T}
\mathbf Q_{nn'}\,
\mathbf q
+
O(\mathbf q^3)
\label{eq:S6}
\end{equation}
The matrix elements $\mathbf P_{nn'}$ and $\mathbf Q_{nn'}$ are given by
\begin{equation}
\mathbf P_{nn'}
=
\int_{\mathrm{u.c.}}
\psi_{n,\mathbf K}^{*}(\mathbf r)
\left[
-i\left(2\rho^{-1}(\mathbf r)\nabla+\nabla \rho^{-1}(\mathbf r)\right)
\right]
\psi_{n',\mathbf K}(\mathbf r)\, d\mathbf r
\label{eq:S7}
\end{equation}
\begin{equation}
\mathbf Q^{(ij)}_{nn'}
=
2\delta_{ij}
\int_{\mathrm{u.c.}}
\psi_{n,\mathbf K}^{*}(\mathbf r)\,
\rho^{-1}(\mathbf r)\,
\psi_{n',\mathbf K}(\mathbf r)\,
d\mathbf r
\label{eq:S8}
\end{equation}
Unlike the linear coefficient $\mathbf P_{nn'}$, whose matrix elements are nonzero only when the $n$ and $n'$ bands are of different parities, the second-order coefficient $\mathbf Q^{(ij)}_{nn'}$ depends on the symmetry of the quadratic momentum factor $\mathbf q_i \mathbf q_j$. In particular, the terms $\mathbf q_x^2$ and $\mathbf q_y^2$ are even under both mirror reflections and therefore contribute to same-parity channels, whereas the mixed term $q_x q_y$ is odd under both reflections and can couple states whose parity difference is $(-,-)$.

Next, near the Dirac point, the four relevant $s$, $p_x$, $p_y$ and $d_{xy}$ modes are aimed, we define the basis
\begin{equation}
\Psi=
\bigl(
|s\rangle,\,
|d_{xy}\rangle,\,
|p_x\rangle,\,
|p_y\rangle
\bigr)^T
\label{eq:S9}
\end{equation}

The mirror operations are
\begin{equation}
M_x:(x,y)\rightarrow(-x,y),
\qquad
M_y:(x,y)\rightarrow(x,-y)
\label{eq:S10}
\end{equation}
From the eigenmodes as shown in Figure~\ref{fig:Fig. S2}b), the parity of the four basis states is
\begin{equation}
|s\rangle:(+,+),
\qquad
|d_{xy}\rangle:(-,-),
\qquad
|p_x\rangle:(-,+),
\qquad
|p_y\rangle:(+,-)
\label{eq:S11}
\end{equation}
At the $M$ point, the $p_x$ and $p_y$ modes form a degenerate doublet,
\begin{equation}
\omega_{p_x}(M)=\omega_{p_y}(M)\equiv \omega_p(M)
\label{eq:S12}
\end{equation}
while the $s$- and $d_{xy}$-derived modes are nondegenerate and exchange order across the transition with the changing d, as shown in Figure~\ref{fig:Fig. S2}.

The momentum deviations transform as
\begin{equation}
q_x:(-,+),
\qquad
q_y:(+,-)
\label{eq:S13}
\end{equation}
and therefore
\begin{equation}
q_x^2:(+,+),\qquad
q_y^2:(+,+),\qquad
q_xq_y:(-,-)
\label{eq:S14}
\end{equation}
Accordingly, the symmetry-allowed linear couplings are
\begin{equation}
\langle s|H|p_x\rangle \propto q_x,
\qquad
\langle s|H|p_y\rangle \propto q_y,
\qquad
\langle d_{xy}|H|p_x\rangle \propto q_y,
\qquad
\langle d_{xy}|H|p_y\rangle \propto q_x
\label{eq:S15}
\end{equation}

Thus, the first-order Hamiltonian is
\begin{equation}
H^{(1)}(\mathbf q)=
\begin{pmatrix}
0 & 0 & a q_x & b q_y \\
0 & 0 & c q_y & d q_x \\
a^* q_x & c^* q_y & 0 & 0 \\
b^* q_y & d^* q_x & 0 & 0
\end{pmatrix}
\label{eq:S16}
\end{equation}

At second order, the symmetry-allowed terms include diagonal band-curvature terms $\beta_{ix}q_x^2+\beta_{iy}q_y^2$, $i=s,d_{xy},p_x,p_y$, as well as the off-diagonal couplings $\langle s|H|d_{xy}\rangle = \gamma q_xq_y$, and $\langle p_x|H|p_y\rangle = \eta q_xq_y$.

Therefore, the second-order Hamiltonian is
\begin{equation}
H^{(2)}(\mathbf q)=
\begin{pmatrix}
\beta_{sx}q_x^2+\beta_{sy}q_y^2 & \gamma q_xq_y & 0 & 0 \\
\gamma^* q_xq_y & \beta_{dx}q_x^2+\beta_{dy}q_y^2 & 0 & 0 \\
0 & 0 & \beta_{px}q_x^2+\beta_{py}q_y^2 & \eta q_xq_y \\
0 & 0 & \eta^* q_xq_y & \tilde\beta_{px}q_x^2+\tilde\beta_{py}q_y^2
\end{pmatrix}
\label{eq:S20}
\end{equation}
where $\beta_{ix}$ and $\beta_{iy}$ describe the quadratic band curvature of the $i$-th mode along the $x$ and $y$ directions, respectively, and originate from the diagonal components of the second-order expansion tensor $Q^{(ij)}_{nn'}$; $\gamma$ characterizes the leading-order coupling between the $s$ and $d_{xy}$ modes and arises from the off-diagonal component $Q^{(xy)}_{s,d_{xy}}$, which is proportional to $q_x q_y$, consistent with the symmetry requirement that only operators with parity $(-,-)$ can connect the two modes; $\eta$ characterizes the leading second-order coupling between the degenerate $p_x$ and $p_y$ modes. Similar to the $s$-$d_{xy}$ coupling, the $\eta$ terms are proportional to $q_x q_y$ and arises from the symmetry-allowed off-diagonal component of the second-order expansion tensor.

Combining the zeroth-, first-, and second-order terms, the full effective Hamiltonian up to second order is
\begin{equation}
H(\mathbf q)=
\begin{pmatrix}
\omega_s^2+\beta_{sx}q_x^2+\beta_{sy}q_y^2
&
\gamma q_xq_y
&
a q_x
&
b q_y
\\[6pt]
\gamma^* q_xq_y
&
\omega_d^2+\beta_{dx}q_x^2+\beta_{dy}q_y^2
&
c q_y
&
d q_x
\\[6pt]
a^* q_x
&
c^* q_y
&
\omega_{p_x}^2+\beta_{px}q_x^2+\beta_{py}q_y^2
&
\eta q_xq_y
\\[6pt]
b^* q_y
&
d^* q_x
&
\eta^* q_xq_y
&
\omega_{p_y}^2+\tilde\beta_{px}q_x^2+\tilde\beta_{py}q_y^2
\end{pmatrix}
\label{eq:S21}
\end{equation}

In our study, the unit cells forming the topological insulators are square unites that follows $C_4$ symmetry. As a result, the dispersions along the $k_x$ and $k_y$ directions are symmetry-equivalent, and the Hamiltonian can be simplified using
\begin{equation}
\omega_{p_x}=\omega_{p_y}\equiv \omega_p, \qquad
\text{and} \qquad
\beta_{sx}=\beta_{sy}\equiv \beta_s,\qquad
\beta_{dx}=\beta_{dy}\equiv \beta_d,\qquad
\beta_{px}=\tilde{\beta}_{py}, \qquad 
\beta_{py}=\tilde{\beta}_{px}
\label{eq:S23}
\end{equation}

Exactly at the $M$ point in the first BZ, $\mathbf q=0$, so the Hamiltonian reduces to
\begin{equation}
H(0)=
\mathrm{diag}
\bigl(
\omega_s^2,\,
\omega_d^2,\,
\omega_p^2,\,
\omega_p^2
\bigr)
\label{eq:S24}
\end{equation}
Therefore, the eigenvalues at this high symmetry point are
\begin{equation}
E_s=\omega_s^2,
\qquad
E_d=\omega_d^2,
\qquad
E_{p_x}=E_{p_y}=\omega_p^2
\label{eq:S25}
\end{equation}

This shows that the $p_x$ and $p_y$ modes remain degenerate at $M$, whereas the $s$ and $d_{xy}$ modes are the ones that exchange order.

We define the bulk mass parameter as
\begin{equation}
m_{\mathrm{bulk}}
=
\frac{\omega_d^2-\omega_s^2}{2}
\label{eq:S26}
\end{equation}
where the critical point is reached when $\omega_s=\omega_d$, that is $d=\sqrt2 a_1/4$ in the unit cell, corresponding to the crossing point in Figure~\ref{fig:Fig. S2}a), so that
\begin{equation}
m_{\mathrm{bulk}}=0
\label{eq:S28}
\end{equation}
The sign of $m_{\mathrm{bulk}}$ distinguishes the two phases:
$m_{\mathrm{bulk}}>0$ corresponds to one ordering of $s$ and $d_{xy}$, donating to the trivial lattice, whereas
$m_{\mathrm{bulk}}<0$ corresponds to the inverted ordering, donating to the nontrivial lattice.

Since the topological transition is governed by the $s$-$d_{xy}$ inversion, the low-energy physics can be projected into the subspace
\begin{equation}
(|s\rangle,\ |d_{xy}\rangle)
\label{eq:S34}
\end{equation}
Let $\tau_i$ be the Pauli matrices in this reduced subspace. Owing to the symmetry of the woven lattice, the dispersion near the $M$ point is equivalent along the $k_x$ and $k_y$ directions, so the reduced Hamiltonian can be written as
\begin{equation}
H_{\mathrm{eff}}(\mathbf q)
=
\bar\omega^2 I
+
m(\mathbf q)\tau_z
+
v(q_x\tau_x+q_y\tau_y)
+
\delta H^{(2)}(\mathbf q),
\label{eq:S35}
\end{equation}
where
\begin{equation}
\bar\omega^2=\frac{\omega_s^2+\omega_d^2}{2},
\label{eq:S36_new}
\end{equation}
\begin{equation}
m(\mathbf q)=m_{\mathrm{bulk}}+m_2(q_x^2+q_y^2),
\label{eq:S37}
\end{equation}
and $\delta H^{(2)}(\mathbf q)$ collects the symmetry-allowed quadratic corrections, including the direct $s$-$d_{xy}$ coupling term proportional to $q_xq_y$. The corresponding eigenvalues are
\begin{equation}
E_\pm(\mathbf q)
=
\bar\omega^2
\pm
\sqrt{
m(\mathbf q)^2+v^2(q_x^2+q_y^2)+\mathcal{O}(q^3,q^4)
}.
\label{eq:S38}
\end{equation}
At the critical geometry, $m_{\mathrm{bulk}}=0$, and the bulk gap closes at the $M$ point. Away from this point, the sign of $m_{\mathrm{bulk}}$ distinguishes the trivial and nontrivial phases.


To connect the bulk transition to the corner-localized states of the finite woven crystal, we further characterize the two phases by the bulk polarization
\begin{equation}
\mathbf{P}=(P_x,P_y),
\label{eq:S39}
\end{equation}
with
\begin{equation}
P_i=-\frac{1}{(2\pi)^2}\sum_{n\in \mathrm{occ}}
\int_{\mathrm{BZ}} d^2\mathbf{k}\;
i\left\langle u_n(\mathbf{k}) \middle| \partial_{k_i}u_n(\mathbf{k}) \right\rangle,
\qquad i=x,y,
\label{eq:S40}
\end{equation}
where the summation runs over the bands below the band gap of interest. Equivalently, the Zak phases along the two lattice directions are
\begin{equation}
\theta_i=2\pi P_i,
\qquad i=x,y.
\label{eq:S41}
\end{equation}

In the woven lattice, the critical degeneracy at the $M$ point determines the bulk topological transition through the sign change of $m_{\mathrm{bulk}}$. This transition is accompanied by a change in the bulk polarization, such that the trivial and nontrivial phases are distinguished by

\begin{equation}
\Delta \mathbf{P}
=
\mathbf{P}^{(\mathrm{nontriv})}
-
\mathbf{P}^{(\mathrm{triv})}
\neq 0
\quad (\mathrm{mod}\ 1).
\label{eq:S42}
\end{equation}

When a nontrivial woven region is embedded in the trivial lattice, this polarization mismatch produces edge-localized states along the edge interfaces. For a square finite crystal, the two orthogonal edge interfaces are symmetry-related, and their intersection gives a corner charge:
\begin{equation}
q_{\mathrm{corner}}=(\Delta P_x+\Delta P_y)\bmod 1.
\label{eq:S43}
\end{equation}

A nonzero corner charge, $q_{\mathrm{corner}}$,implies zero-dimensional localization at the corners, which is the higher-order topological response observed in the finite woven structure. In this way, the corner states are understood as the finite-structure manifestation of the bulk topological transition: the $M$-point $k\cdot p$ theory captures the critical band inversion, while the change in bulk polarization establishes the bulk–edge–corner correspondence. Their existence is confirmed directly by the in-gap corner branches in the finite-supercell spectrum and by the corner-localized eigenmodes.

\medskip
\textbf{Acknowledgements} \par 
C.M. and M.W acknowledge the funding support from NSF CAREER award - Communications, Circuits, and Sensing Systems (CCSS) Program [\#: 2237619] and Wisconsin Alumni Research Foundation. P.P. and H.T. acknowledge the funding support from NSF CAREER award - Mechanics of Materials and Structures (MOMS) Program [\#: 2046476].

\medskip
\textbf{Conflict of Interest} \par
The authors declare no conflict of interest.

\medskip 
\textbf{Data Availability Statement} \par
The data that support the findings of this study are available from the corresponding author upon reasonable request.

\medskip 
\bibliographystyle{MSP}
 \bibliography{reference.bib}

\end{document}